%% file: main.tex
\documentclass[10pt, conference]{IEEEtran}
\IEEEoverridecommandlockouts
\usepackage{cite}
\usepackage{enumitem}
\usepackage{amsmath,amssymb,amsfonts}
\usepackage{algorithmic}
\usepackage{graphicx}
\usepackage{textcomp}
\usepackage[dvipsnames]{xcolor}
\usepackage{xspace} 
\usepackage{threeparttable}
\usepackage{booktabs}
\usepackage{multirow}
\usepackage{pdfpages}
\usepackage{graphicx}
\usepackage{amsmath}
\usepackage{balance}

\usepackage[colorlinks=true, urlcolor=blue, linkcolor=blue, bookmarks=false]{hyperref}

\usepackage[a4paper, total={184mm,239mm}]{geometry}
\def\BibTeX{{\rm B\kern-.05em{\sc i\kern-.025em b}\kern-.08em
    T\kern-.1667em\lower.7ex\hbox{E}\kern-.125emX}}

\newcommand{\ours}{\emph{VeriDebug}}
\newcommand{\vpara}[1]{\noindent\textbf{#1}\xspace}
\begin{document}

\title{VeriDebug: A Unified LLM for Verilog Debugging via Contrastive Embedding and Guided Correction}

\author{
% First row - authors
Ning Wang$^{1}$ \quad Bingkun Yao$^{1}$ \quad Jie Zhou$^{2}$ \quad Yuchen Hu$^{2}$ \quad Xi Wang$^{2}$ \quad Nan Guan$^{1}$ \quad Zhe Jiang$^{2}$ \\[0.5em]
% Second row - affiliations
$^{1}$City University of Hong Kong \quad $^{2}$Southeast University
}

\maketitle

\begin{abstract}

Large Language Models (LLMs) have demonstrated remarkable potential in debugging for various programming languages. However, the application of LLMs to Verilog debugging remains insufficiently explored. 
% \guan{debugging comes suddenly; why bother debugging?}
% \guan{before telling what techniques are used, first motivate why it is challenging and what challenges we are addressing}.
Here, we present \ours{}, an approach that integrates contrastive representation and guided correction capabilities for automated Verilog debugging.
Unlike existing methods, \ours{} employs an embedding-based technique to accurately retrieve internal information, followed by bug-fixing. 
\ours{} unifies Verilog bug detection and correction through a shared parameter space. By simultaneously learning bug patterns and fixes, it streamlines debugging via contrastive embedding and guided correction. 
Empirical results show the efficacy of \ours{} in enhancing Verilog debugging. 
Our \ours{}$_{\text{Loc, Type}}$  model achieves 64.7\% accuracy in bug fixing ($Acc@1$), a significant improvement from the existing open-source SOTA's 11.3\%. 
This performance not only outperforms open-source alternatives but also exceeds larger closed-source models like GPT-3.5-turbo (36.6\%), offering a more accurate alternative to conventional debugging methods.
The Repo. of dataset and code: \url{https://anonymous.4open.science/r/VeriDebug-F770/}.

\end{abstract}

%\begin{IEEEkeywords}
%Verilog Debugging, Large Language Model, Constrastive Information Retrieval, Guided Bug Correction
% \end{IEEEkeywords}

\vspace{-5pt}
\section{Introduction}\label{s:intro}
\vspace{-5pt}

% Large Language Models (LLMs) have revolutionized natural language processing. Their unprecedented scale and complexity enable them to capture intricate language patterns, generating human-like text across diverse topics. LLMs are increasingly applied to various domains, including programming languages \cite{chen2021evaluating}. These models, trained on vast corpora of text and code, demonstrate remarkable capabilities in understanding, generating, and manipulating code across different programming languages \cite{chen2023debug}. The application of LLMs to hardware description languages, particularly Verilog, represents a significant advancement in electronic design automation (EDA).

Large Language Models (LLMs) have revolutionized natural language processing, enabling the generation of human-like text across diverse topics due to their unprecedented scale and complexity. Their application to programming languages has demonstrated remarkable capabilities in understanding, generating, and manipulating code across different languages \cite{chen2021evaluating, chen2023debug}. Specifically, the integration of LLMs into hardware description languages like Verilog marks a significant advancement in Electronic Design Automation (EDA).

% Recent research by \cite{wang2023evaluating, wang2023debug} has explored LLMs' potential in Verilog code generation, proposing a benchmarking framework to evaluate their performance in hardware design contexts. 
% However, in real-world scenarios, debugging and fixing existing Verilog code is often more prevalent and time-consuming than generating new code from scratch \cite{lahti2022synthesis}.
% This practical consideration underscores the importance of developing robust methods for automated Verilog debugging using LLMs. The challenge of Verilog debugging has been approached from various angles. Some researchers have employed Retrieval-Augmented Generation (RAG) techniques, leveraging external information to enhance the debugging process \cite{tsai2023debug}. While effective, these methods may introduce dependencies on external knowledge bases that might not always be available or up-to-date. Other approaches have focused on generating internal information through LLMs, but these methods can potentially suffer from hallucination issues, where the model produces plausible but incorrect information \cite{chen2023debug}.

However, in real-world scenarios, debugging existing Verilog code is often more prevalent and time-consuming than generating new code from scratch \cite{lahti2022synthesis}. Despite this practical necessity, current research has predominantly focused on leveraging Large Language Models (LLMs) to fix bugs through generative methods. These approaches involve the LLM generating corrected versions of the code, effectively "rewriting" faulty sections. While generative techniques have shown promise in automating code fixes, they are susceptible to hallucination issues—where the model produces plausible but incorrect code—which can undermine the reliability of the debugging process \cite{chen2023debug}.

With that, researchers have integrated guidance mechanisms into the generative process. Techniques like Retrieval-Augmented Generation (RAG) incorporate external information to enhance the LLM's outputs during debugging tasks \cite{tsai2023debug}. However, RAG methods introduce their own set of challenges. Relying on external knowledge bases can be problematic if these resources are outdated, inaccessible, or infeasible to use in certain environments. Also, the models used for retrieval often depend on surface-level lexical similarity rather than deep semantic understanding, which may result in retrieving irrelevant or wrong information, leading to misguided fixes and compromise the effectiveness of the debugging process.

% To address these challenges, we propose \ours{}, a novel approach that integrates specialized representation and generation capabilities for automated hardware description language debugging. Our method leverages embedding tasks to retrieve internal information, followed by a bug-fixing phase. This approach aims to mitigate the limitations of existing methods by reducing reliance on external information and minimizing the risk of hallucination.
The key challenge, hence, lies in developing robust automated Verilog debugging methods that effectively harness the generative capabilities of LLMs while solving these limitations. 
It's essential to enhance the reliability and correctness of bug fixes by guiding the generative process with accurate and relevant information, without undue dependence on external resources or the risk of brining new errors through hallucinations.

% \zhenote{A key problem here is, what is the real challenge? We introduce some related work and its problems. However, the reviewer/audience might to see the \textbf{***root and common***} causes of them? If we just solve the rpobelme of [1][2], that is our homework. If we can solve a root and common problem, it will be a good paper. }
% \zhenote{Following above, if we do not know the real challange, it is unclear what kind of problem have been solved.}

% \ours{} employs distinct instructions for various tasks within the representation and generation phases. The representation phase utilizes embedding instructions to process input and generate vector representations, while the generation phase leverages these representations along with bug information and design specifications to produce corrected Verilog code. By streamlining Verilog debugging through the automation of both bug detection and correction processes, \ours{} facilitates simultaneous learning of bug information retrieval and corrected code generation.
\noindent \textbf{Contributions.}
Here, we propose \ours{}, a framework that directly tackles these issues by employing guided generation techniques. 
Our method enhances the LLM's ability to debug Verilog code through a two-phase approach: representation and generation. In the representation phase, \ours{} utilizes embedding instructions to process input and generate vector representations for both the code and relevant information. The generation phase then leverages information retrieved based on the similarity between instruction embeddings and these representations, along with extracted bug information and design specifications. This approach reduces reliance on potentially unreliable external knowledge bases and minimizes the risk of hallucinations by providing the generative model with precisely retrieved, context-relevant guidance. By incorporating open-source models, \ours{} ensures transparency and adaptability, fostering community-driven improvements. By addressing the root causes of deficiencies in existing approaches and streamlining the debugging process through effective information retrieval and guided code correction, \ours{} significantly improves the effectiveness and reliability of automated debugging in hardware design contexts. Our contributions represent a substantial advancement in applying LLMs to Verilog debugging, offering promising implications for the broader field of EDA.
The main contributions of this paper are:

\begin{itemize}[leftmargin=1.5em, itemsep=0pt, parsep=0.2em, topsep=0.1em, partopsep=0.0em]
    \item \ours{}: A debug framework with open-source models, ensuring transparency and fostering collaboration.
    \item An embedding-based method to reduce reliance on external knowledge and mitigate hallucination risks.
    \item An open-source dataset of 8,000 Verilog snippets with bugs and corrections to support further research.
    \item Extensive experiments demonstrating \ours{}'s effectiveness in Verilog debugging across various bug types.
\end{itemize}

\begin{figure*}[!t]
\centering
    \vspace{-35pt}
    \includegraphics[width=\textwidth]{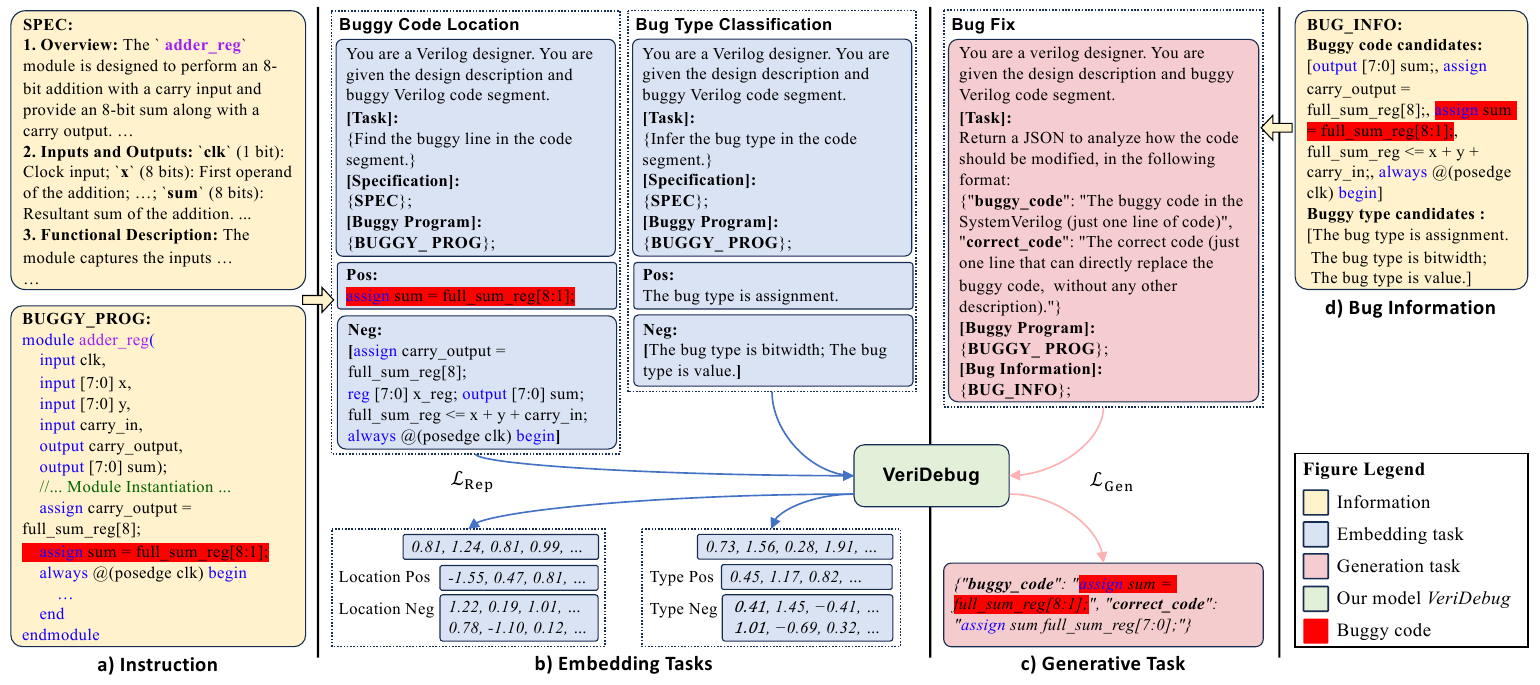}
    \caption{
    % \textbf{Overview of \ours{} for Verilog bug detection and correction.}
    % The representation task employs embedding instructions to transform Verilog code, specifications, and buggy information into vector representations. The generation task uses these representations along with additional instructions to produce corrected Verilog code. The figure also showcases example inputs and outputs for both tasks, demonstrating the flow from buggy code to corrected code.
    \textbf{Overview of \ours{} for Verilog bug detection and correction.}
    This figure shows VeriDebug's architecture and training methodology for identifying and correcting Verilog code bugs. The system learns to analyze buggy inputs and produce corrected outputs using embedding techniques and generative modeling.
    \textbf{a) Instruction:} Containing \texttt{SPEC} (specification) and \texttt{BUGGY\_PROG} (buggy program) to provide essential context for analysis.
    \textbf{b) Embedding Tasks:} Employing contrastive learning for bug location and classification to transform inputs into vector representations. This distinguishes between positive (actual bugs) and negative examples.
    \textbf{c) Generative Task:} Producing corrected Verilog code using information retrieved based on similarity between instruction and candidate information to Outputbuggy code and its correction.
    \textbf{d) Bug Information:} Providing extra context about potential bug locations and types, used as input for the generative bug fix process. In training, we use a mix of ground truth and randomly sampled candidates for bug lines and types, simulating real-world scenarios.
    }
        \vspace{-15pt}

    % \zhenote{If Fig.1 is only referred only once in the paper, why do we need it? A simple solution is add circled a, b, c for each block of the paper.}
\label{fig:overview}
\end{figure*}

\vpara{Note:} 
% However, in hardware design, protecting intellectual property is crucial, making closed-source LLMs problematic due to their security risks when processing proprietary designs, particularly in complex System-on-Chip (SoC) environments \cite{saha2024, alsaqer2023}. Recent research has highlighted unique security and privacy challenges associated with LLMs across various scenarios, including pre-training, fine-tuning, and deployment \cite{liu2024}. Consequently, there is a pressing need for secure, open-source solutions that can be deployed locally, ensuring the confidentiality and integrity of hardware designs while leveraging LLMs' potential for generating sustainable and efficient hardware.
Our paper utilizes open-source models to address key challenges in hardware design automation. Protecting intellectual property in hardware design is crucial, making closed-source LLMs problematic due to security risks when processing proprietary designs, particularly in complex SoC environments \cite{saha2024, alsaqer2023}. Recent research has highlighted unique security and privacy challenges with LLMs in various scenarios \cite{liu2024}. Consequently, there is a pressing need for secure, open-source solutions deployable locally, ensuring confidentiality and integrity of hardware designs while leveraging LLMs' potential for generating efficient hardware.

% By addressing the specific challenges of Verilog debugging, we aim to demonstrate the potential of LLMs in enhancing the efficiency and accuracy of hardware design processes, ultimately contributing to the advancement of EDA tools and methodologies.
% \zhenote{The above sentences mean nothing to me, how it integrate into the EDA tools, an add-on? And what do the experimental results show to us?}

% \input{figs_with_caption/overview}
\vspace{-3pt}
\section{Background}
\vspace{-4pt}
\subsection{Embedding or Generative Models}
\vspace{-3pt}

The field of Natural Language Processing (NLP) has witnessed remarkable both embedding and generative-based applications. Embedding models, such as text-embedding-ada-002, focus on creating dense vector representations of text that capture semantic relationships \cite{neel2022embed}. Recent work has explored multilingual embeddings, enabling zero-shot cross-lingual applications \cite{huang2022multilingual}. Generative language models, exemplified by GPT families, excel at producing human-like text across diverse tasks \cite{openai2022gpt}. These models have shown remarkable few-shot and zero-shot learning capabilities, pushing the boundaries of natural language understanding and generation. 
Recent research also explored the integration of embedding and generative capabilities within unified models. Concept embedding generation \cite{lake2024concept} represents a promising direction, potentially enabling learning new concepts on-the-fly. GRIT \cite{fu2024grit} represents a significant advancement in unifying generative and embedding tasks. 
%By training model to handle both types of tasks through instruction-based differentiation, GRIT achieves state-of-the-art performance on both embedding and generative tasks.

\vspace{-4pt}
\subsection{Verilog Code Debugging with LLMs}
\vspace{-3pt}

In recent years, LLMs have demonstrated remarkable capabilities in understanding, generating, and debugging code across multiple programming languages \cite{thakur2023verigen}. This trend has extended to hardware description languages (HDLs) like Verilog. Open-source models are particularly important in this context as they offer transparency, accessibility, and the ability to customize solutions for specific needs, fostering innovation and collaboration within the community. Code debugging with LLMs involves utilizing the model's vast knowledge to identify and rectify errors in source code. For Verilog, this approach is particularly promising due to the language's complexity and the critical nature of hardware design \cite{verilogeval2024}. LLMs can analyze Verilog code, detect syntax errors, logical inconsistencies, and potential design flaws, offering suggestions for improvement.

\section{Methodology}
\vspace{-3pt}
\subsection{Overview}
\vspace{-4pt}

Our approach, \ours{}, addresses the core challenges of automated hardware description language debugging by integrating specialized representation and generation capabilities, as illustrated in Figure \ref{fig:overview}. This figure provides a comprehensive overview of \ours{}'s architecture and training methodology for Verilog bug detection and correction. The system learns to analyze buggy inputs and produce corrected outputs using embedding techniques and generative modeling.

The architecture of \ours{} begins with the Instruction component (Figure \ref{fig:overview}a), which encapsulates \texttt{SPEC} (specification) and \texttt{BUGGY\_PROG} (buggy program). This initial element provides essential context for code implementation, serving as the primary input for the system's analytical processes.

The representation phase of \ours{} employs embedding techniques that process inputs to generate vector representations of queries, positive responses, and negative responses, as depicted in Figure \ref{fig:overview}b. This approach benefits the debugging process by creating a rich, context-aware internal knowledge base, addressing the problem of dependency on potentially outdated or inaccessible external resources. By leveraging contrastive learning, our model develops a deep semantic understanding of Verilog code, overcoming the limitations of surface-level lexical similarity often observed in conventional retrieval models.

The generation phase of \ours{}, illustrated in Figure \ref{fig:overview}c, utilizes extracted information guided by embeddings, rather than directly using the representations from the previous phase. This process involves using cosine similarity to retrieve relevant information from the buggy code and design specifications, which are provided as part of the instruction input (Figure \ref{fig:overview}a). By leveraging this extracted, relevant data, the generation phase produces corrected Verilog code. This approach addresses the critical issue of hallucination by grounding the generation process in accurate, contextually appropriate information obtained through embedding-based similarity matching.

By combining these representation and generation capabilities, \ours{} achieves unique features such as simultaneous learning of bug information retrieval and corrected code generation. This integration allows for efficient capture of intricate relationships between bug patterns and their corresponding corrections within a shared parameter space. These features collectively address the fundamental challenges in RTL debugging, offering a more reliable and context-aware solution. By reducing dependency on external resources and minimizing hallucination risks, \ours{} significantly improves the accuracy of automated Verilog debugging. The use of open-source models is crucial in this process, as they provide transparency, flexibility, and community-driven improvements, ensuring that the solutions remain adaptable and robust. While this approach may require additional computational resources, it prioritizes the quality and reliability of bug fixes, which is crucial in the complex domain of hardware description language debugging. This focus on enhanced accuracy and reduced error rates in bug detection and correction ultimately contributes to more robust and dependable RTL debug processes.

\vspace{-4pt}
\subsection{Joint Training}
\vspace{-3pt}

\ours{} integrates representation learning for bug information retrieval and generative learning for bug correction into a unified model. This approach fine-tunes a pre-trained large language model using both bug location and bug correction instruction data in a consistent format.

For buggy code location and bug type classification, \ours{} employs a contrastive objective to compute the loss, represented by the equation:

\footnotesize
\vspace{-3pt}
\[
\mathcal{L}_{\text{Rep}} = - \frac{1}{M} \sum_{i=1}^M \log \frac{\exp(\tau \cdot \sigma(f_\theta(q^{(i)}), f_\theta(d^{(i)})))}{\sum_{j=1}^M \exp(\tau \cdot \sigma(f_\theta(q^{(i)}), f_\theta(d^{(j)})))}
\]
\vspace{-3pt}
\normalsize 

It aims to maximize the similarity between matched query-document pairs while minimizing similarity with non-matching pairs. 
Here, $f_\theta$ represents the model parameterized by $\theta$, while $\tau$ is the temperature parameter controlling the distribution's softness. The symbol $\sigma$ denotes the pooling operation followed by cosine similarity computation. Query and document samples are represented by $q$ and $d$ respectively, with $M$ denoting the batch size. The index $i$ iterates over the batch, selecting query-document pairs $(q^{(i)}, d^{(i)})$, while the index $j$ in the denominator compares each query $q^{(i)}$ against all documents $d^{(j)}$, creating positive pairs when $i=j$ and negative pairs when $i \neq j$.

For bug correction, \ours{} uses the following objective:

\footnotesize
\[
\vspace{-3pt}
\mathcal{L}_{\text{Gen}} = - \sum_{t=1}^T \log P_\pi(\hat{y}_t | \mathbf{x}, \hat{\mathbf{y}}_{<t})
\vspace{-3pt}
\]
\normalsize

This function maximizes the probability of generating the correct token sequence for bug correction.

The two objectives are combined with a weight $\lambda$:

\footnotesize
\[
\vspace{-3pt}
L = \mathcal{L}_{\text{Rep}} + \lambda\mathcal{L}_{\text{Gen}}
\vspace{-3pt}
\]
\normalsize

This combined loss function enables \ours{} to simultaneously learn bug information retrieval and corrected code generation, capturing intricate relationships between bug patterns and their corrections within a shared parameter space.

\vspace{-4pt}
\subsection{Cascade Inference}
\vspace{-3pt}

The inference process of \ours{}, as illustrated in Fig.~\ref{fig:infer}, presents a structured approach to debugging Verilog code. The process begins with \ours{} taking two essential inputs: the specification (\texttt{SPEC}) and the buggy program (\texttt{BUGGY\_PROG}). These inputs form the foundation for the debugging procedure.

\ours{} then performs embedding on the embedding tasks and their candidates, transforming these elements into representation vectors for a more comprehensive analysis. Simultaneously, it conducts bug type classification as part of this embedding process. The figure depicts these embeddings as sequences of numerical values (e.g., 0.81, 1.24, 0.81, 0.99, \ldots). This integrated approach allows \ours{} to capture both the structural information of the code and potential bug characteristics in a unified representation.

Following the embedding and classification, the model applies a similarity-based ranking, calculating cosine similarity between the query and candidate embeddings. 
This step, which utilizes the contrastive learning from the training, is crucial for identifying the most relevant information for debugging.

With the ranking, \ours{} generates \texttt{BUG\_INFO}, e.g., identifying buggy code candidates. The combined embedding and classification process from earlier enables \ours{} to efficiently pinpoint the location and nature of the bugs.

With this comprehensive bug information at hand, \ours{} processes a "Buggy Fix" task, analyzing the detected issues and their classified types to determine appropriate corrections. The final output of this process is the generation of corrected Verilog code in JSON format. This output specifies both the buggy code and the correct code, offering a clear comparison. As illustrated in the figure, it identifies the buggy code \texttt{assign sum = full\_sum\_reg[8:1];} and generates the correction \texttt{assign sum = full\_sum\_reg[7:0];}.

\input{figs_with_caption/infer}

\section{A New Synthetic Dataset}

A key contribution of our paper is addressing the critical shortage of high-quality, diverse data for training models in Verilog debugging tasks. To solve this problem, we propose a novel method for generating synthetic data that includes both buggy code and fixed code. As detailed in Table \ref{tab:dataset}, our synthetic dataset encompasses a diverse range of bug types commonly encountered in Verilog code. This approach not only provides ample training data but also ensures a rich, varied dataset that closely mimics real-world Verilog debugging scenarios.

Our synthetic data generation process leverages GPT-4 to create a comprehensive and realistic dataset of buggy Verilog code. The process involves three main steps:

\input{tables/dataset}

\begin{itemize}[leftmargin=1.5em, itemsep=0pt, parsep=0.2em, topsep=0.1em, partopsep=0.0em]
\item \textbf{Data Preparation:} We initiate the process with a corpus of valid Verilog code, meticulously ensuring that all modules are complete and syntactically correct. The original dataset is sourced from \cite{thakur2023verigen}. To maintain data integrity, we employ Icarus Verilog to conduct a syntax check, filtering out any non-compliant code samples.
\item \textbf{Bug Injection:} GPT-4 performs a thorough analysis of each Verilog module, generating a diverse list of potential bugs that could realistically occur in the code. These identified bug types serve as a guide for GPT-4 to insert errors into the code. The bug categories encompass a range of common Verilog coding errors, including width, logic, assignment, initial, data, state, comparison, bitwise, signal, arithmetic, and value-related issues. The injected bugs are deliberately designed to be subtle, accurately representing the nuanced errors that often occur in real-world Verilog development.
\item \textbf{Verification:} To ensure the dataset's focus on functional bugs rather than syntactical errors, we subject the modified code to compilation using Icarus Verilog. This crucial step verifies that the injected bugs do not induce compilation errors, maintaining the dataset's integrity and relevance to practical debugging scenarios.
\end{itemize}

The resulting dataset comprises approximately $8,000$ samples, each containing a detailed bug description, the modified (buggy) line of code, and the original correct line. This pairing of buggy and correct code provides essential context for model training in bug detection and correction tasks.
We construct an assessment dataset using stratified sampling, selecting $10\%$ of the Buggy Code samples across various code length intervals. This approach ensures a representative distribution of modules for robust model evaluation.
This dataset is open-source as a part of the open-source project of this work.

%Open-source datasets are crucial for hardware companies as they enable collaborative development, foster innovation, and ensure that advancements in debugging technologies are accessible to a broader community. Our method provides a large, diverse, and realistic dataset that enables more effective training and evaluation of debugging models, potentially advancing the state-of-the-art in automated Verilog debugging and repair.

% \zhenote{This is a true problem. Tesla does not have enough real-world data for their automated cars, therefore, they explored the why to generate synthetic data.}

% \subsection{Tasks}

% In this paper, we incorporates two embedding tasks and one generative task for Verilog bug information retrieval and fixing. 

% % Table \ref{tab:task} delineates the instructions and positive and negative responses (if embedding) for different tasks.

% The first embedding task, line location, segments the code into individual lines and applies filtering to identify potential bug locations. The second embedding task, classification, employs a detailed categorization scheme to determine the specific type of bug present in the code segment.

% The generative task focuses on code correction. It requires the production of a JSON object that specifies the necessary modifications to the code. This object includes two key elements: the buggy code (a single line of SystemVerilog) and the correct code (a single line that can directly replace the buggy code). 

% \zhenote{What questions do these experiments try to answer? Correctness? Effectiveness of each step? }

\section{Experiments \& Evaluation}

\subsection{Training Techniques}

% We use VeriSeek \cite{wang2023debug} as our base model and fine-tune all parameters, employing a learning rate of 2e-5 with a linear warm-up phase encompassing 3\% of the total steps, followed by a linear decay to zero. To manage computational resources efficiently, the methodology incorporates several advanced techniques, including PyTorch Fully Sharded Data Parallel (FSDP), gradient checkpointing, BF16 mixed precision training, and Flash-Attention 2 via PyTorch Scaled Dot-Product Attention (SDPA).

% The number of negative samples in an embedding instance is 7 and we set the balance hyper-parameter $\lambda$ as 1. The training phase utilizes varying sequence lengths: 2048 tokens for generative samples, 256 for embedding queries, and 2048 for embedding documents, unless specified otherwise. The Adam optimizer is employed with $\beta_1=0.9$ and $\beta_2=0.999$, without weight decay. The whole fine-tuning process requires approximately 8 hours, utilizing a cluster of eight high-performance 80GB A100 GPU.

To rigorously evaluate our method's accuracy, we conducted comprehensive experiments using VeriSeek \cite{wang2023debug} as our base model. We fine-tuned all parameters with a learning rate of 2e{-5}, employing a linear warm-up for the initial 3\% of steps, followed by a linear decay to zero, optimizing model convergence and accuracy.

To ensure reliable and efficient accuracy-focused experiments, we implemented advanced computational techniques:

\begin{itemize}[leftmargin=1.5em, itemsep=0pt, parsep=0.2em, topsep=0.1em, partopsep=0.0em]
    \item PyTorch Fully Sharded Data Parallel (FSDP): Enabling distributed training to scale experiments and potentially improve accuracy through larger batch sizes.
    \item Gradient checkpointing: Reducing memory to train on longer sequences, getting more context for better accuracy.
    \item BF16 mixed precision training: Maintaining numerical stability and accelerating computations for more experiments.
    \item Flash-Attention 2 via PyTorch Scaled Dot-Product Attention (SDPA): Optimizing attention computations to train on larger datasets and capture nuanced patterns.
\end{itemize}

Our embedding strategy used 7 negative samples per instance, with the balance hyper-parameter $\lambda$ set to 1. We employed varying sequence lengths: 2048 tokens for generative samples, 256 for embedding queries, and 2048 for embedding documents, to assess accuracy across different input lengths.

We used the Adam optimizer ($\beta_1=0.9, \beta_2=0.999$, no weight decay) for stable and accurate training. The fine-tuning, aimed at maximizing accuracy, took about 8 hours on eight A100 GPUs (80GB), underscoring our commitment to thoroughly validating our approach in generating Verlog.

\begin{table}[]
\centering
\begin{threeparttable}
\setlength{\tabcolsep}{4pt}
\renewcommand\arraystretch{1.2}
\caption{Comparison of open-source baselines and proposed Language Models' Performance on Verilog Debugging Tasks}
\vspace{-5pt}
\begin{tabular}{p{1cm}p{2.0cm}p{0.5cm}p{0.6cm}p{0.6cm}p{0.6cm}p{0.6cm}p{0.6cm}}
\toprule
\multirow{3}{*}{\textbf{Type}}          & \multirow{3}{*}{\textbf{Model}} & \multicolumn{5}{c}{\textbf{Embedding Tasks}}                                         & \textbf{Gen} \\  
\cline{3-8} &  & \multicolumn{3}{c}{Location} & \multicolumn{2}{c}{Type} & Fix         \\  
 & & $A@1$ & $A@3$ & $A@5$ & $A@1$ & $A@3$ & $A@1$ \\
\specialrule{.08em}{.2em}{.2em} 
                      
% \multirow{2}{*}{\textbf{Closed-source}} & GPT-3.5-turbo          & 40.4      & 44.7      & 63.7     & 29.0               & 41.4               & 36.8            \\
%                       & GPT-4o                   & 77.9    & 90.1    & 93.9    & 57.6    & 88.7    & 75.0    \\ \midrule
\multirow{2}{*}{\textbf{Baselines}}   & Deepseek-coder         & 12.1      & 28.3      & 42.8     & 7.1                & 18.8               & 8.5             \\
                      & VeriSeek                 & 15.2    & 32.7    & 45.5    & 7.9     & 19.4    & 11.3    \\ \midrule
\multirow{3}{*}{\textbf{Ours}} & \ours{}                & 39.5    & 47.2    & 50.7    & 8.8     & 21.7    & 36.6    \\
                      & \ours{}$_{\text{Loc}}$       & 66.5    & 80.9    & 86.2    & 9.4     & 23.0    & 60.6    \\
                      & \ours{}$_{\text{Loc, Type}}$ & \textbf{67.2}    & \textbf{81.5}    & \textbf{86.9}    & \textbf{48.6}    & \textbf{79.5}    & \textbf{64.7}   \\
\bottomrule
\end{tabular}
\begin{tablenotes} 
    \item[*] All models in this table have approximately 7 billion parameters.
    \item[+] Bold font represents the best metric. 
    % \guan{should/can find some better excuse for excluding GPT-4? For example, in reality people won't use closed-source models such as GPT-4 as they don't want to leak their design information?}
    \item[$\dagger$] \textbf{Gen} represents generation tasks and $A$ represents $Acc$.
\vspace{-10pt}
\end{tablenotes}
\label{tab:rslt}
\end{threeparttable}
\end{table}

\subsection{Metrics}
We employ top-$k$ accuracy metrics to assess our large language model's performance across various code-related tasks. The general formula for top-$k$ accuracy is:

\footnotesize
\[
    Acc@k = \frac{\text{Correct predictions in top $k$}}{\text{Total predictions}} \times 100\%
\]
\normalsize

\subsubsection{Location Task}: $Acc@1$, $Acc@3$, and $Acc@5$ evaluate code location predictions.
\subsubsection{Type Classification Task}: $Acc@1$ and $Acc@3$ measure bug type classification performance.
\subsubsection{Bug Fix Task}: $Acc@1$ with strict criterion - only exact matches of fixed code line to label are correct.

\begin{figure}[t]
\centering
\vspace{-10pt}
\includegraphics[width=0.48\textwidth]{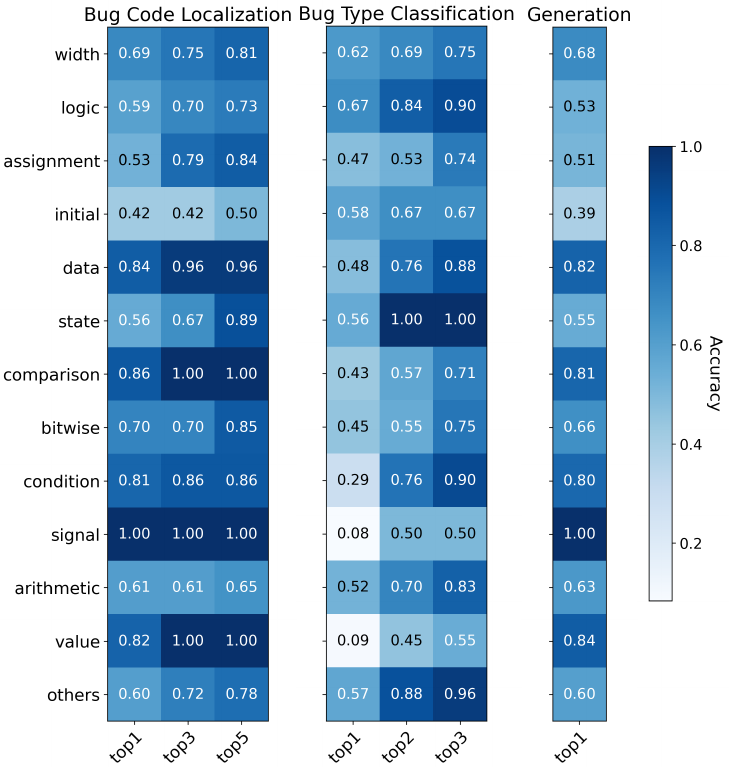}
\vspace{-8pt}
\caption{
    \textbf{Detailed results of \ours{} on different bug types.} 
    % remove close-source models;
    }
\vspace{-15pt}
\label{fig:detail}
\end{figure}

% \zhenote{As we discussed, please add more experiments.}
\subsection{Results}
% Table \ref{tab:rslt} compares the performance of various language models on Verilog debugging tasks, including Bug Location, Type Classification, and Bug Fix. The models are categorized into three groups: Closed-source (GPT-3.5-turbo and GPT-4), Open-source (Deepseek-coder and VeriSeek), and our proposed model with its variants. Our model is presented in three forms: \ours{} (trained only with the generative task), \ours{}$_{\text{Loc}}$ (incorporating bug location embedding and generative tasks), and \ours{}$_{\text{Loc, Type}}$ (combining bug location embedding, type classification embedding, and generative tasks). To facilitate comparison, the table uses bold font to highlight the best metric scores among all models, excluding GPT-4, which allows for a more balanced evaluation among models with similar resource requirements and availability.
Table \ref{tab:rslt} compares the performance of 
our proposed model (with different configurations) with Deepseek-coder \cite{deepseekcoder}, the SOTA open-source coding-oriented LLM and VeriSeek \cite{wang2023debug}, a recently published open-source LLM for verilog coding. Our model is presented in three configurations: \ours{} (trained exclusively on the generative task), \ours{}${\text{Loc}}$ (integrating bug location embedding with generative tasks), and \ours{}${\text{Loc, Type}}$ (combining bug location embedding, type classification embedding, and generative tasks). 

% The main conclusion from Table \ref{tab:rslt} is that our proposed model, particularly the \ours{}$_\text{Loc,Type}$ variant, achieves state-of-the-art (SOTA) debugging capacity among open-source models and outperforms GPT-3.5-turbo. This demonstrates the effectiveness of our proposed methods in Verilog language debugging tasks. Specifically, in Bug Location, our best model achieves 67.2\%, 81.5\%, and 86.9\% for $Acc@1, Acc@3, and Acc@5$ respectively, significantly outperforming other open-source models and GPT-3.5-turbo. In Type Classification, our model reaches 48.6\% and 79.5\% for $Acc@1$ and $Acc@3$, again surpassing other open-source models and GPT-3.5-turbo. In Bug Fix, our model achieves 64.2\% $Acc@1$, which is substantially higher than other open-source models and even outperforms GPT-3.5-turbo (36.8\%).
Table \ref{tab:rslt} shows that our model, particularly the \ours{}$_{\text{Loc, Type}}$ variant, outperforms its competitors. 
Table \ref{tab:rslt} also reveals additional insights into the performance characteristics of our model variants. The bug location accuracy ($Acc@1$ for the Bug Location) closely approximates the bug fix accuracy ($Acc@1$ for the Bug Fix) for our models. For instance, \ours{}$_{\text{Loc, Type}}$ achieves 67.2\% $Acc@1$ for Bug Location and 64.2\% $Acc@1$ for Bug Fix. This correlation indicates that precise bug localization is critical for successful Verilog debugging, highlighting the importance of our multi-task approach.

Furthermore, the incorporation of embedding tasks markedly enhances performance across all metrics. The bug location embedding task (\ours{}$_{\text{Loc}}$) significantly improves performance compared to the base model (\ours{}). Building upon this, the additional inclusion of the type classification embedding task (\ours{}$_{\text{Loc, Type}}$) yields further improvements, particularly in Type Classification and Bug Fix tasks. However, it is worth noting that the enhancement from adding type information is less pronounced than that from adding location information, suggesting that detailed location information is more vital for Verilog debugging than type information.

Beyond comparisons with other open-source models, our proposed model also demonstrates competitive performance against prominent closed-source alternatives. Notably, it outperforms GPT-3.5-turbo across key metrics. In Bug Code Location, our \ours{}$_{\text{Loc, Type}}$ model achieves 67.2\%, 81.5\%, 86.9\% versus GPT-3.5-turbo's 40.4\%, 44.7\%, 63.7\% for $Acc@1, Acc@3, Acc@5$. For Type Classification, we attain 48.6\%, 79.5\% versus GPT-3.5-turbo's 40.4\% for $Acc@1, Acc@3$. The most significant difference is observed in the Bug Fix task, where our model reaches 64.2\% $Acc@1$, far surpassing GPT-3.5-turbo's 36.8\%.
While GPT-4 can achieve 77.9\%, 90.1\% 93.9\% for bug location $Acc@1, Acc3, Acc@5$; 57.6\%, 88.7\% for bug location $Acc@1, Acc3$; and 75.0\% for final bug fix $Acc@1$.
Note that the comparison with GPT-3.5-turbo and GPT-4 is only for reader's reference. As discussed in Section \ref{s:intro}, different from software development, in hardware design domain developers are unlikely willing to use closed-source LLMs for many reasons, like IP protection and security.
 
%our results demonstrate the considerable potential of open-source models in specialized tasks like Verilog debugging. This progress is particularly valuable for hardware companies preferring open-source solutions, suggesting that competitive performance can be achieved without relying on closed-source alternatives. Our research thus paves the way for further advancements in open-source language models tailored for specific programming languages and debugging tasks, potentially revolutionizing the landscape of hardware development tools.

\begin{figure}[t]
\centering
\vspace{-10pt}
\includegraphics[width=0.45\textwidth]{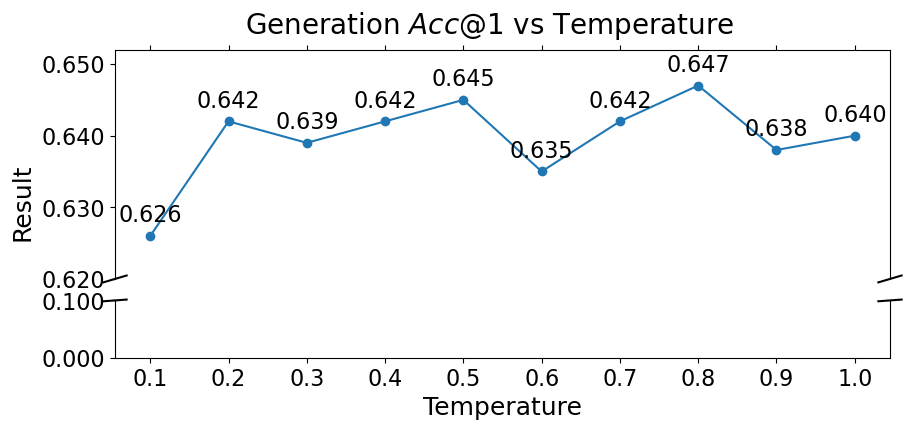}
\vspace{-10pt}
\caption{
    \textbf{Generation $Acc@1$ under different temperature.} 
    % remove close-source models;
    }
\vspace{-15pt}
\label{fig:acc_temp}
\end{figure}

\subsection{Detailed Analysis}
Figure \ref{fig:detail} illustrates the performance of a bug detection system across three tasks: Bug Code Localization, Bug Type Classification, and Generation for various bug types.
In Bug Code Localization, \texttt{signal} bugs achieve perfect accuracy (1.00) across all metrics, while \texttt{initial} bugs are the most challenging. The Bug Type Classification task reveals unexpected results: \texttt{state} bugs excel, but \texttt{signal} and \texttt{value} bugs, easily localized, are difficult to classify (top1 accuracy of 0.08 and 0.09). For Generation, \texttt{signal} bugs again score perfectly (1.00), with \texttt{initial} bugs remaining most challenging (0.39).
This suggests that a bug type's difficulty varies across different tasks, suggesting the need for tailored strategies. Notably, \texttt{initial} bugs consistently challenge the system due to their contextual complexity, subtle nature, and diverse manifestations. These bugs often involve incorrect initialization, with effects that may only become apparent later in execution, making them difficult to detect, classify, and generate across all tasks.
This analysis underscores the complex nature of bug detection and the importance of developing versatile, multi-faceted approaches.

Figure \ref{fig:acc_temp} illustrates our model's robustness in bug fix accuracy (Acc@1) across temperatures ranging from 0.1 to 1.0. The consistent performance observed throughout these temperature variations underscores the model's stability. This resilience suggests that the guided information effectively directs the model's generation process, enabling it to maintain reliable bug-fixing capabilities regardless of temperature adjustments. 
%The model's ability to leverage this guidance contributes to its consistent accuracy across different randomness levels in token selection.

\subsection{Hallucination in Debug Process}

Figure \ref{fig:hallucination} illustrates hallucination in LLM debugging. Instruction (Figure \ref{fig:hallucination}a) shows the actual buggy code \texttt{Clock12C <= Clock12C;} in the \texttt{BaudRate12C} module. Figure \ref{fig:hallucination}b demonstrates hallucination by incorrectly identifying non-existent code \texttt{Clock12C <= !Clock12C;} as the bug with base model \ours{}. In contrast as shown in Figure \ref{fig:hallucination}c, \ours{}$_{\text Loc}$ accurately locates and fixes the real bug with \texttt{Clock12C <= ~Clock12C;}. This comparison highlights how location-aware debugging prevents hallucinations and provides accurate fixes, unlike naive approaches that may invent non-existent issues.

\input{figs_with_caption/hallucination}

\subsection{Ablation Study}

We conducted an ablation study to evaluate the efficacy of various methodological components in addressing the three tasks. This comparative analysis, as illustrated in Table~\ref{tab:performance_comparison}, elucidates the performance implications of different approaches, including Context Candidates, Negative Samples, and Unidirectional Attention, in comparison to our proposed method. All performance metrics are reported as $Acc@1$, representing the accuracy of the top-1 prediction.

The Context Candidates approach, which involves considering the surrounding code rather than isolated lines, yielded suboptimal results across all tasks. With $Acc@1$ scores of 34.6\%, 25.2\%, and 32.7\% for Bug Location, Type Classification, and Bug Fix respectively, this method significantly underperformed compared to our proposed approach. These results suggest that the inclusion of contextual information may have introduced noise, potentially obfuscating the salient features necessary for accurate task performance.

The implementation of additional negative samples in contrastive learning, specifically employing 15 negative samples, showed moderate improvement over the Context Candidates method. This approach achieved $Acc@1$ scores of 59.0\%, 41.4\%, and 53.1\% across the three tasks. However, it is lower than our method, indicating that while the inclusion of additional data provided some benefit, it did not fully capture the complexity of the tasks.

The Unidirectional Attention method, which employs a left-to-right attention mechanism in the generation of embeddings, demonstrated the best performance among the baseline approaches. With $Acc@1$ scores of 62.6\% for Bug Location, 44.3\% for Type Classification, and 58.8\% for Bug Fix, this method came closest to matching the performance of our proposed approach. The effectiveness of directional attention in capturing relevant features for these software engineering tasks is evident, yet there remained room for improvement.

Our proposed method consistently outperformed all other approaches across all three tasks. It achieved the highest $Acc@1$ scores of 67.2\% for Bug Location, 48.6\% for Type Classification, and 64.2\% for Bug Fix. Compared to the best-performing baseline (Unidirectional Attention), our method showed improvements of $4.6\%$, $4.3\%$, and $5.4\%$ in the three tasks, respectively.

\section{Discussions}
\textbf{Connection with RAG}.
While both \ours{} and RAG leverage embedding techniques for information retrieval, their approaches diverge significantly in the context of Verilog code analysis \cite{lewis2020retrieval}. \ours{} represents a paradigm shift in how we approach hardware design verification, moving beyond the traditional RAG framework to address the challenges of Verilog code analysis and debug.
The key distinction lies in the nature of the information being retrieved and processed. Unlike RAG, which typically relies on external knowledge bases, \ours{} focuses on extracting and utilizing internal bug-related information directly from the Verilog code \cite{wang2023retrieval}. This introspective approach allows for a more nuanced understanding of potential hardware design flaws, as it operates within the specific context of the code rather than relying on generalized external knowledge.
Unlike traditional RAG systems that may sometimes struggle with fully understanding the context of a query, \ours{} leverages a shared parameter space for both bug detection and correction in Verilog code. This enables our model to capture the intricate relationships between bug patterns and their corresponding corrections. 
%It is a crucial capability which not only enhances the accuracy of bug identification but also leads to more coherent and appropriate fix suggestions, ultimately improving the efficiency and reliability of the Verilog code verification process.

\begin{table}[t]
\centering
\caption{Ablation study: Performance comparison across different software engineering tasks. The table shows the performance metrics for various methods on Bug Location, Type Classification, and Bug Fix tasks.}
\begin{tabular}{lcccc}
\toprule
\multirow{2}{*}{Method} & \multicolumn{3}{c}{Task Performance} \\
\cmidrule(lr){2-4}
 & Location & Classification & Fix \\
\midrule
Context Candidates & 34.6 & 25.2 & 32.7 \\
Negative samples (15) & 59.0 & 41.4 & 53.1 \\
Unidirectional Attention & 62.6 & 44.3 & 58.8 \\
\textbf{Ours} & \textbf{67.2} & \textbf{48.6} & \textbf{64.2} \\
\bottomrule
\end{tabular}
\label{tab:performance_comparison}
\end{table}

\textbf{Potential Scaling Law for Embedding Tasks}.
The scaling law phenomenon have demonstrated the importance in improving model performance \cite{kaplan2022scaling}. As models grow in size and complexity, their performance on a wide range of natural language processing tasks improves in a predictable manner. 
\cite{shao2024scaling} has shown that scaling retrieval-based large language models can yield performance improvements comparable to those observed in traditional LLMs. Their work on scaling retrieval-based language models with a trillion-token datastore demonstrates the potential for significant performance gains through increased scale. This finding lends credence to the hypothesis that scaling embedding tasks could potentially enhance bug detection and resolution capabilities.
However, it is crucial to acknowledge the potential drawbacks of indiscriminate scaling. While increasing the number of embedding tasks may improve performance, it could also introduce noise and potentially lead to diminished results. The optimal balance between task quantity and quality remains an open question that warrants further investigation.

\section{Conclusion}

Our research introduces \ours{}, a method for Verilog debugging using Large Language Models. By integrating contrastive representation and guided correction, \ours{} achieves significant improvements in bug detection and correction. Our \ours{}$_{\text{Loc, Type}}$ model demonstrates 64.7\% accuracy in bug fixing ($Acc@1$), outperforming both open-source alternatives and larger closed-source models like GPT-3.5-turbo. This work not only advances automated debugging techniques for hardware description languages but also showcases the potential of LLMs in this crucial domain.

% \section*{Acknowledgment}

\bibliographystyle{IEEEtran}
\balance

\end{document}

%% file: figs_with_caption/infer.tex
\begin{figure}[t]
\centering
\vspace{-5pt}
\includegraphics[width=0.47\textwidth]{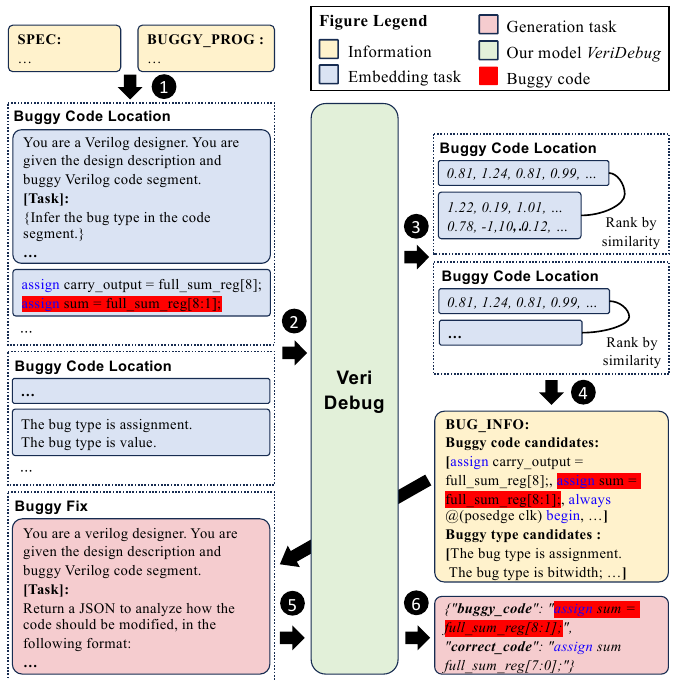}
\vspace{-10pt}
\caption{
    \textbf{Inference process of \ours{}:} 
    \ours{} is a model designed to debug Verilog code by identifying and correcting bugs. 
    \includegraphics[page=1,width=1em]{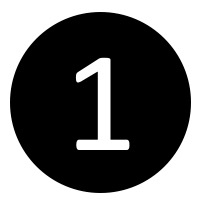} The process begins with the model taking \texttt{SPEC} and \texttt{BUGGY\_PROG} as input. 
    \includegraphics[page=1,width=1em]{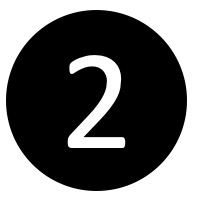} It then performs embedding on the buggy code location. 
    \includegraphics[page=1,width=1em]{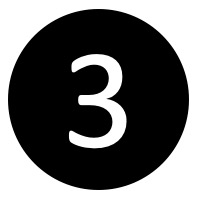} Similarity-based ranking is applied to identify relevant embeddings. 
    \includegraphics[page=1,width=1em]{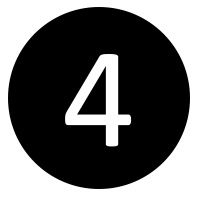} \texttt{BUG\_INFO} is generated, including buggy code candidates and bug type candidates. 
    \includegraphics[page=1,width=1em]{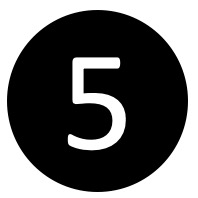} The model processes a "Buggy Fix" task. 
    \includegraphics[page=1,width=1em]{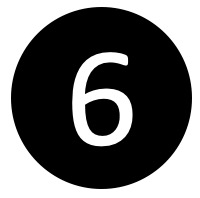} Finally, it generates the corrected Verilog code in JSON format, specifying the buggy code and the correct code.
    }
\vspace{-15pt}
\label{fig:infer}
\end{figure}

%% file: tables/dataset.tex
\begin{table*}[]
\centering
\vspace{-35pt}
\caption{Verilog bug types and examples. \textbf{Bold} fonts represents bug code and \textcolor{blue}{Blue} font represents keywords for simple read.}
\vspace{-5pt}
 \resizebox{0.99\linewidth}{!}{%
\begin{tabular}{p{1.2cm}p{12cm}p{3.5cm}p{3.5cm}}
\toprule
\textbf{Bug Type} &
\textbf{Description} &
\textbf{Original} &
\textbf{Bug} \\
\specialrule{.08em}{.2em}{.2em}
Width &
Mismatched bit widths causing unintended truncation / zero-extension &
\textcolor{blue}{input} \textbf{{[}6:0{]}} data\_offset\_delay; &
\textcolor{blue}{input} \textbf{{[}5:0{]}} data\_offset\_delay; \\
\midrule
Logic &
Errors in combinational / sequential logic causing wrong behavior / timing &
\textcolor{blue}{assign} rnw \textbf{=} writeback\_w; &
\textcolor{blue}{assign} rnw \textbf{=$\sim$} writeback\_w; \\
\midrule
Assignment &
Improper use of blocking/non-blocking assignments causing race conditions/unexpected updates &
int\_dout\_next = \textbf{1'b0}; &
int\_dout\_next = \textbf{1'bx}; \\ \midrule
Initial &
Incorrect initialization of variables/registers causing undefined behavior/simulation. &
e\_tx\_ack $\text{\texttt{\scriptsize{<=}}}$ \textbf{1'b0}; &
e\_tx\_ack $\text{\texttt{\scriptsize{<=}}}$ \textbf{1'b1}; 
\\ \midrule
Data &
Errors in data handling, including incorrect types/improper conversions/misuse of signed/unsigned values &
wb\_sel\_o $\text{\texttt{\scriptsize{<=}}}$ \#1 \textbf{cbu\_sel\_i}; &
wb\_sel\_o $\text{\texttt{\scriptsize{<=}}}$ \#1 \textbf{4'b0000}; \\ \midrule
State &
Flaws in FSM design, including missing states/incorrect transitions/improper encoding &
state $\text{\texttt{\scriptsize{<=}}}$ \textbf{STATE\_INIT}; &
state $\text{\texttt{\scriptsize{<=}}}$ \textbf{STATE\_START}; \\ \midrule
Comparison &
Incorrect use of equality/inequality operators or misuse of case equality/inequality. &
counter \textbf{==} 7 &
counter \textbf{!=} 7  \\ \midrule
Bitwise &
Errors in bitwise operations, including incorrect use of AND, OR, XOR, or shift operators. &
PCplusout $\text{\texttt{\scriptsize{<=}}}$ PCplusin; &
PCplusout $\text{\texttt{\scriptsize{<=}}}$ PCplusin \textbf{$\textgreater{}\textgreater$ 1}; \\ \midrule
Condition &
Flaws in conditional statements (if-else, case) leading to incorrect branching or priority encoding issues. &
\textcolor{blue}{if} (\textbf{pause==0}) &
\textcolor{blue}{if} (\textbf{pause==1}) \\ \midrule
Signal &
Errors in signal declarations, including incorrect wire/reg use, input/output ports/naming conflicts &
\textcolor{blue}{assign} Out = \textbf{Switch}; &
\textcolor{blue}{assign} Out = \textbf{Out1}; \\ \midrule
Arithmetic &
Mistakes in arithmetic operations, including overflow/underflow issues/incorrect signed/unsigned usage &
ndCount $\text{\texttt{\scriptsize{<=}}}$ count \textbf{+} limit; &
ndCount $\text{\texttt{\scriptsize{<=}}}$ count \textbf{-} limit; \\ \midrule
Value &
Incorrect constant values/parameter definitions/literal representations causing unexpected behavior &
\textcolor{blue}{localparam} OPCODE = \textbf{6'h04}; &
\textcolor{blue}{localparam} OPCODE = \textbf{6'h05}; \\ \midrule
Others &
Miscellaneous errors that don't fit into other categories, such as syntax errors or tool-specific issues. &
\textbf{supply1} vddio; &
\textbf{supply0} vddio; \\ 
\bottomrule
\end{tabular}
}
\label{tab:dataset}
\vspace{-12pt}
\end{table*}

%% file: figs_with_caption/hallucination.tex
\begin{figure}[htbp]
\centering
\includegraphics[width=0.5\textwidth]{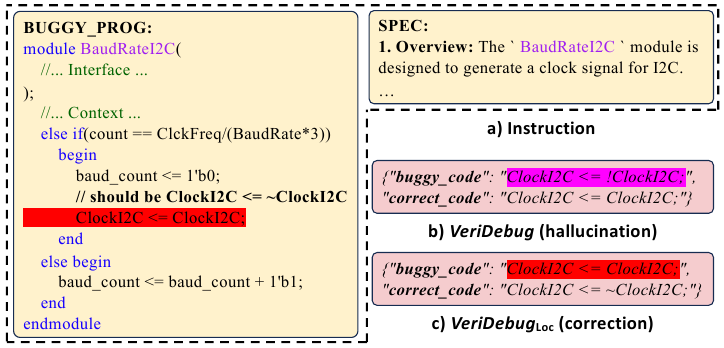}
\vspace{-20pt}
\caption{
    \textbf{Hallucination example.} \textcolor{red}{Red} color represents the buggy code while \textcolor{VioletRed}{violet} shows the hallucination code which is not available in \texttt{BUGGY\_PROG}.
} 
\vspace{-10pt}
\label{fig:hallucination}
\end{figure}

%% file: main.bbl
\begin{thebibliography}{00}

\bibitem{saha2024}
Saha, D., Tarek, S., Yahyaei, K., Saha, S. K., Zhou, J., Tehranipoor, M., \& Farahmandi, F. (2024). Llm for soc security: A paradigm shift. \textit{IEEE Access}.

\bibitem{alsaqer2023}
Alsaqer, S., Alajmi, S., Ahmad, I., \& Alfailakawi, M. (2024). The potential of LLMs in hardware design. \textit{Journal of Engineering Research}.

\bibitem{liu2024}
Wang, S., Zhu, T., Liu, B., Ming, D., Guo, X., Ye, D., \& Zhou, W. (2024). Unique Security and Privacy Threats of Large Language Model: A Comprehensive Survey. \textit{arXiv preprint arXiv:2406.07973}.

\bibitem{lahti2022synthesis}
Lahti, S., Sjövall, P., Vanne, J., \& Hämäläinen, T. D. (2018). Are we there yet? A study on the state of high-level synthesis. \textit{IEEE Transactions on Computer-Aided Design of Integrated Circuits and Systems, 38}(5), 898-911.

\bibitem{tsai2023debug}
Tsai, Y., Liu, M., \& Ren, H. (2023). Rtlfixer: Automatically fixing rtl syntax errors with large language models. \textit{arXiv preprint arXiv:2311.16543}.

\bibitem{chen2021evaluating}
Chen, M., Tworek, J., Jun, H., Yuan, Q., Pinto, H. P. D. O., Kaplan, J., ... \& Zaremba, W. (2021). Evaluating large language models trained on code. \textit{arXiv preprint arXiv:2107.03374}.

\bibitem{wang2023debug}
Wang, N., Yao, B., Zhou, J., Wang, X., Jiang, Z., 
\& Guan, N. (2024). Large Language Model for Verilog Generation with Golden Code Feedback. \textit{arXiv preprint arXiv:2407.18271}.

\bibitem{deepseekcoder}
Guo, D., Zhu, Q., Yang, D., Xie, Z., Dong, K., Zhang, W., ... \& Liang, W. (2024). DeepSeek-Coder: When the Large Language Model Meets Programming--The Rise of Code Intelligence. arXiv preprint arXiv:2401.14196.

\bibitem{chen2023debug}
Zhong, L., Wang, Z., \& Shang, J. (2024). Ldb: A large language model debugger via verifying runtime execution step-by-step. \textit{arXiv preprint arXiv:2402.16906}.

\bibitem{wang2023evaluating}
Liu, M., Pinckney, N., Khailany, B., \& Ren, H. (2023, October). Verilogeval: Evaluating large language models for verilog code generation. In 2023 IEEE/ACM International Conference on Computer Aided Design (ICCAD) (pp. 1-8). IEEE.

\bibitem{neel2022embed}
Neelakantan, A., Xu, T., Puri, R., Radford, A., Han, J. M., Tworek, J., ... \& Weng, L. (2022). Text and code embeddings by contrastive pre-training. \textit{arXiv preprint arXiv:2201.10005}.

\bibitem{huang2022multilingual}
Kuan-Hao Huang, I-Hung Hsu, Prem Natarajan, Kai-Wei Chang, and Nanyun Peng. 2022. Multilingual Generative Language Models for Zero-Shot Cross-Lingual Event Argument Extraction. In \textit{Proceedings of the 60th Annual Meeting of the Association for Computational Linguistics (Volume 1: Long Papers)}, pages 4633–4646, Dublin, Ireland. Association for Computational Linguistics.

\bibitem{openai2022gpt}
Radford, A., Wu, J., Child, R., Luan, D., Amodei, D., \& Sutskever, I. (2019). Language models are unsupervised multitask learners. \textit{OpenAI blog, 1}(8), 9.

\bibitem{lake2024concept}
Teehan, R., Lake, B., \& Ren, M. (2024). CoLLEGe: Concept Embedding Generation for Large Language Models. \textit{arXiv preprint arXiv:2403.15362}.

\bibitem{thakur2023verigen} 
Thakur, S., Ahmad, B., Fan, Z., Pearce, H., Tan, B., Karri, R., ... \& Garg, S. (2023, April). Benchmarking large language models for automated verilog rtl code generation. In \textit{2023 Design, Automation \& Test in Europe Conference \& Exhibition (DATE)} (pp. 1-6). IEEE.

\bibitem{verilogeval2024}
Liu, M., Pinckney, N., Khailany, B., \& Ren, H. (2023, October). Verilogeval: Evaluating large language models for verilog code generation. In \textit{2023 IEEE/ACM International Conference on Computer Aided Design (ICCAD)} (pp. 1-8). IEEE.

\bibitem{fu2024grit}
Muennighoff, N., Su, H., Wang, L., Yang, N., Wei, F., Yu, T., ... \& Kiela, D. (2024). Generative representational instruction tuning. \textit{arXiv preprint arXiv:2402.09906}.

\bibitem{lewis2020retrieval}
Lewis, P., Perez, E., Piktus, A., Petroni, F., Karpukhin, V., Goyal, N., ... \& Kiela, D. (2020). Retrieval-augmented generation for knowledge-intensive nlp tasks. \textit{Advances in Neural Information Processing Systems, 33}, 9459-9474.

\bibitem{wang2023retrieval}
Gao, Y., Xiong, Y., Gao, X., Jia, K., Pan, J., Bi, Y., ... \& Wang, H. (2023). Retrieval-augmented generation for large language models: A survey. \textit{arXiv preprint arXiv:2312.10997}.

\bibitem{kaplan2022scaling}
Kaplan, J., McCandlish, S., Henighan, T., Brown, T. B., Chess, B., Child, R., ... \& Amodei, D. (2020). Scaling laws for neural language models. arXiv preprint arXiv:2001.08361.

\bibitem{shao2024scaling}
Shao, R., He, J., Asai, A., Shi, W., Dettmers, T., Min, S., ... \& Koh, P. W. (2024). Scaling Retrieval-Based Language Models with a Trillion-Token Datastore. \textit{arXiv preprint arXiv:2407.12854}.

\end{thebibliography}
